\documentclass[10pt,conference]{IEEEtran}

\usepackage{amsfonts}
\usepackage{adjustbox}

\usepackage{float}
\usepackage{algorithm}
\usepackage[noend]{algpseudocode}

\usepackage{amsmath}
\usepackage{graphicx}
\usepackage{caption}

\usepackage{subcaption}

\usepackage{array}
\usepackage{tabularx}
\usepackage{color}
\usepackage{epsf}
\usepackage{times}
\usepackage{epsfig}
\usepackage{epstopdf}
\usepackage{hyperref}
\usepackage{cite}
\usepackage{amssymb}
\usepackage{amsxtra}
\usepackage{amsthm}
\usepackage{bbm}
\usepackage{enumitem}
\usepackage{tikz}
\usepackage{pgfplotstable}
\usepackage{rotate}

\definecolor{rubblue}{cmyk}{1,0.5,0,0.6}
\definecolor{rubgreen}{cmyk}{0.5,0,1,0}
\definecolor{rubgray}{cmyk}{0.03,0.03,0.03,0.1}

\usepgfplotslibrary{units}

\usetikzlibrary{%
patterns,%
calc,%
fit,%
arrows,%
plotmarks,%
shadows,%
chains,%
shapes%
}

\tikzset{>=latex'} 
\tikzstyle{every picture}+=[remember picture] 
\pgfdeclarelayer{background}
\pgfdeclarelayer{foreground}
\pgfsetlayers{background,main,foreground}

\tikzstyle{blueblock}=[draw=rubblue, rectangle, thick, drop shadow, minimum width=20mm, minimum height=8mm,fill=rubblue!20, text width=20mm, text centered]
\tikzstyle{bluebox}=[draw=rubblue, rectangle, thick, drop shadow, minimum width=8mm, minimum height=8mm,fill=rubblue!20, text width=8mm, text centered]
\tikzstyle{greenblock}=[draw=rubgreen, rectangle, thick, drop shadow, minimum width=20mm, minimum height=8mm,fill=rubgreen!20, text width=20mm, text centered]
\tikzstyle{dot} = [draw, circle, minimum size=0.2pt,scale=0.3,fill=black,black]
\tikzstyle{smalldot} = [draw, circle, minimum size=0.1pt,scale=0.2,fill=black,black]
\tikzstyle{reddot}  =[draw,circle,minimum size=0.2pt,scale=0.8,fill=red,thin]
\tikzstyle{greendot}  =[draw,circle,minimum size=0.2pt,scale=0.8,fill=Green,thin]
\tikzstyle{bluedot}  =[draw,circle,minimum size=0.2pt,scale=0.8,fill=blue,thin]
\tikzstyle{whitedot}=[draw,circle,minimum size=0.2pt,scale=0.8,fill=white,thin]
\tikzstyle{blackdot} = [draw, circle, minimum size=0.2pt,scale=0.7,fill=black,black]
\tikzstyle{sum} = [drop shadow, draw=rubblue, thick, fill=rubblue!20, circle]
\tikzstyle{relay} = [blueblock, minimum width=5mm, minimum height=20mm, text width=5mm, rounded corners=2pt]
\tikzstyle{relay2} = [blueblock, minimum width=5mm, minimum height=15mm, text width=5mm, rounded corners=2pt]
\tikzstyle{relay3} = [blueblock, minimum width=5mm, minimum height=25mm, text width=5mm, rounded corners=2pt]
\tikzstyle{relay4} = [blueblock, minimum width=5mm, minimum height=10mm, text width=5mm, rounded corners=2pt]
\tikzstyle{relay5} = [blueblock, minimum width=5mm, minimum height=50mm, text width=5mm, rounded corners=2pt]
\tikzstyle{relay6} = [blueblock, minimum width=5mm, minimum height=5mm, text width=5mm, rounded corners=2pt]
\tikzstyle{circgreen} = [draw, circle, inner sep=2pt, fill=rubgreen, drop shadow, thick]
\tikzstyle{circwhite} = [draw, circle, inner sep=2pt, fill=white, drop shadow, thick]
\tikzstyle{circdashed} = [draw, dashed, circle, inner sep=2pt, fill=rubgray, drop shadow, thick]
\tikzstyle{vertbox} = [rectangle, draw=rubblue, thick, rotate=90, text centered, minimum width=16.5mm, minimum height=8mm, text width=16.5mm, inner sep=0pt, fill=rubblue!20, drop shadow]
\tikzstyle{vertboxb} = [rectangle, draw=rubblue, thick, rotate=90, text centered, minimum width=16.5mm, minimum height=8mm, text width=16.5mm, fill=rubblue!20, drop shadow]
\tikzstyle{vertboxshort} = [rectangle, draw=rubblue, thick, rotate=90, text centered, minimum width=10mm, minimum height=8mm, text width=10mm, inner sep=0pt, fill=rubblue!20, drop shadow]
\tikzstyle{smalldotgreen} = [draw=rubgreen, circle, minimum size=0.2pt,scale=0.8,fill=rubgreen!20]
\tikzstyle{antenna} = [regular polygon, regular polygon sides=3, draw, shape border rotate=180, minimum size=0.2pt, scale=0.3]

\tikzstyle{poly} = [regular polygon, regular polygon sides=6, shape aspect=0.5, minimum width=1.5cm, minimum height=0.35cm, draw, dashed]

\definecolor{cff9e00}{RGB}{255,158,0}
\definecolor{c4fff00}{RGB}{79,255,0}
\definecolor{cff0012}{RGB}{255,0,18}
\definecolor{c00c5ff}{RGB}{0,197,255}
\definecolor{c046f00}{RGB}{4,111,0}
\definecolor{c004b9d}{RGB}{0,75,157}

\newlength{\mylen}
\usetikzlibrary{petri}
\usetikzlibrary{shapes}
\usetikzlibrary{positioning,arrows,patterns}
\usepackage{scalefnt}
\usetikzlibrary{calc,decorations.markings}
\pgfplotsset{compat=1.10}
\usetikzlibrary{arrows.meta}
\usepackage[columnwise,switch,mathlines]{lineno} 

\usepackage{pgfplots}
\pgfplotsset{compat=newest}
\usepgfplotslibrary{groupplots}

\newtheorem{Theorem}{Theorem}
\newtheorem{Lemma}{Lemma}
\IEEEoverridecommandlockouts\IEEEpubid{\makebox[\columnwidth]{ 978-1-6654-3540-6/22~\copyright~2023 IEEE \hfill} \hspace{\columnsep}\makebox[\columnwidth]{ }}

\begin{document}

\title{Exploiting Hybrid Terrestrial/LEO Satellite\\ Systems for Rural Connectivity\vspace{-0.5cm}}


\author{
     \IEEEauthorblockN{
     Houcem Ben Salem\IEEEauthorrefmark{1}\IEEEauthorrefmark{2}, 
    Nour Kouzayha\IEEEauthorrefmark{2}, 
     Ammar EL Falou\IEEEauthorrefmark{2},
     Mohamed-Slim Alouini\IEEEauthorrefmark{2}, 
     and Tareq Y. Al-Naffouri\IEEEauthorrefmark{2}}
     \IEEEauthorblockA{\IEEEauthorrefmark{1}CNR-Istituto di Elettronica e di Ingegneria dell’Informazione e delle Telecomunicazioni (CNR-IEIIT), 10129 Torino, Italy}
     \IEEEauthorblockA{\IEEEauthorrefmark{2}CEMSE Division, King Abdullah University of Science and Technology, Thuwal, 23955, Saudi Arabia}
    \IEEEauthorblockA{Email: houcem.bensalem@ieiit.cnr.it,\{nour.kouzayha, ammar.falou, slim.alouini, tareq.alnaffouri\}@kaust.edu.sa}
     \vspace{-0.8cm}
 }
 
\maketitle

\begin{abstract}
Satellite networks are playing an important role in realizing global seamless connectivity in beyond 5G and 6G wireless networks. In this paper, we develop a comprehensive analytical framework to assess the performance of hybrid terrestrial/satellite networks in providing rural connectivity. We assume that the terrestrial base stations are equipped with multiple-input-multiple-output (MIMO) technologies and that the user has the option to associate with a base station or a satellite to be served. Using tools from stochastic geometry, we derive tractable expressions for the coverage probability and average data rate and prove the accuracy of the derived expressions through Monte Carlo simulations. The obtained results capture the impact of the satellite constellation size, the terrestrial base station density, and the MIMO configuration parameters. 
\end{abstract}

\vspace{-0.2cm}
\section{Introduction}
With the ever-evolving demands of universal coverage and high-quality communications, the integration of terrestrial networks and space networks is recognized as a key pillar of the sixth-generation (6G) of wireless systems~\cite{dang2020should}. Such integrated networks can also help in overcoming the digital divide in rural areas where deploying more terrestrial base stations is costly and infeasible~\cite{yaacoub2020key}. Terrestrial cellular networks are usually deployed to cover suburban, urban, and dense areas, typically having a positive return on investment (ROI). However, in rural environments, the ROI becomes negative, and terrestrial network operators tend to deploy base stations (BSs) only in villages or main roads~\cite{Falou}. Users around these locations can be covered by the few available cellular BSs and/or by other non-terrestrial networks such as satellite constellations~\cite{kodheli2020satellite}.

Among space networks, low earth orbit (LEO) satellite constellations are gaining importance and are massively deployed worldwide~\cite{yue2022security}. LEO networks are more suitable than other types of satellite solutions as geostationary and medium orbit ones, thanks to their low latency and large capacity~\cite{kodheli2020satellite}. However, unlike terrestrial BSs, satellites are usually more expensive and need a long deployment time. Thus, network planning for hybrid terrestrial/satellite networks is mandatory to sustain cost-effective deployment and respect the 
operator targets while serving efficiently users in rural areas. The implementation of multiple-input-multiple-output (MIMO) technologies ~\cite{khaled2021multi} in the terrestrial network is one way to increase the perceived data rate per user and the overall system capacity of the hybrid terrestrial/satellite network~\cite{arsal2021coverage,falou2023exploring}. In this work, we consider a hybrid terrestrial/LEO satellite system and study its performance using tools from stochastic geometry. We focus on a rural environment, typically having lower densities of terrestrial BSs. We also assume that the terrestrial BSs are equipped with multiple antennas. 

Tools from stochastic geometry have been widely used to analyze the performance of terrestrial and aerial networks~\cite{kouzayha2021analysis,kouzayha2020stochastic}. These tools have been recently extended to characterize satellite networks~\cite{wang2022ultra,okati2022nonhomogeneous,Sun}. For instance, the authors in~\cite{al2021analytic,Park} study the coverage probability of a LEO satellite system and capture the effect of the number and height of satellites and the effective isotropic radiated power per satellite. Furthermore, the authors in~\cite{Homssi,Song} consider hybrid terrestrial/satellite networks and provide analytical relations between the coverage performance and the system design parameters. Using stochastic geometry, the authors in~\cite{park2023unified} explore the benefits of satellites in extending the coverage of terrestrial networks and offloading data in dense urban areas. Finally, the work in~\cite{Lim} aims at modeling the co-channel interference and out-of-band leakage power from mmWave terrestrial networks to satellite networks. To sum up, none of the existing works on modeling hybrid terrestrial/satellite networks have studied an extreme rural scenario, which is the main focus of this work. Furthermore, the existing works have not considered the impact of mounting the terrestrial BSs with multiple antennas to serve users in rural areas. 

In this work, we develop a tractable mathematical framework using stochastic geometry to evaluate the performance of downlink (DL) hybrid terrestrial/satellite networks in offering rural connectivity. The terrestrial BSs are sparsely distributed and are equipped with MIMO antennas to increase the data rate per user and the overall system capacity.
To the best of the authors' knowledge, this is the first work that considers a hybrid terrestrial/satellite network while accounting for the impact of the MIMO antennas of the terrestrial network in a rural environment. The main contributions of this work can be summarized as
\begin{itemize}[noitemsep,nolistsep]
    \item Development of a tractable framework for studying the performance of DL hybrid terrestrial/satellite networks in rural areas.
    \item Derivation of analytical expressions for the main performance metrics, including the association probability with each network, the coverage probability, and the average achievable data rate.
    \item Extraction of analytical relations between the derived performance metrics and the operational parameters for both networks, including; the number of satellites and transmit power, the BS density, and the MIMO configuration. The numerical results clearly highlight the importance of hybrid terrestrial/LEO satellite systems in offering connectivity in rural areas. 
\end{itemize}
 
\section{System model}\label{sec:system}
\begin{figure}[t!]
\centering
   \includegraphics[width=\linewidth]{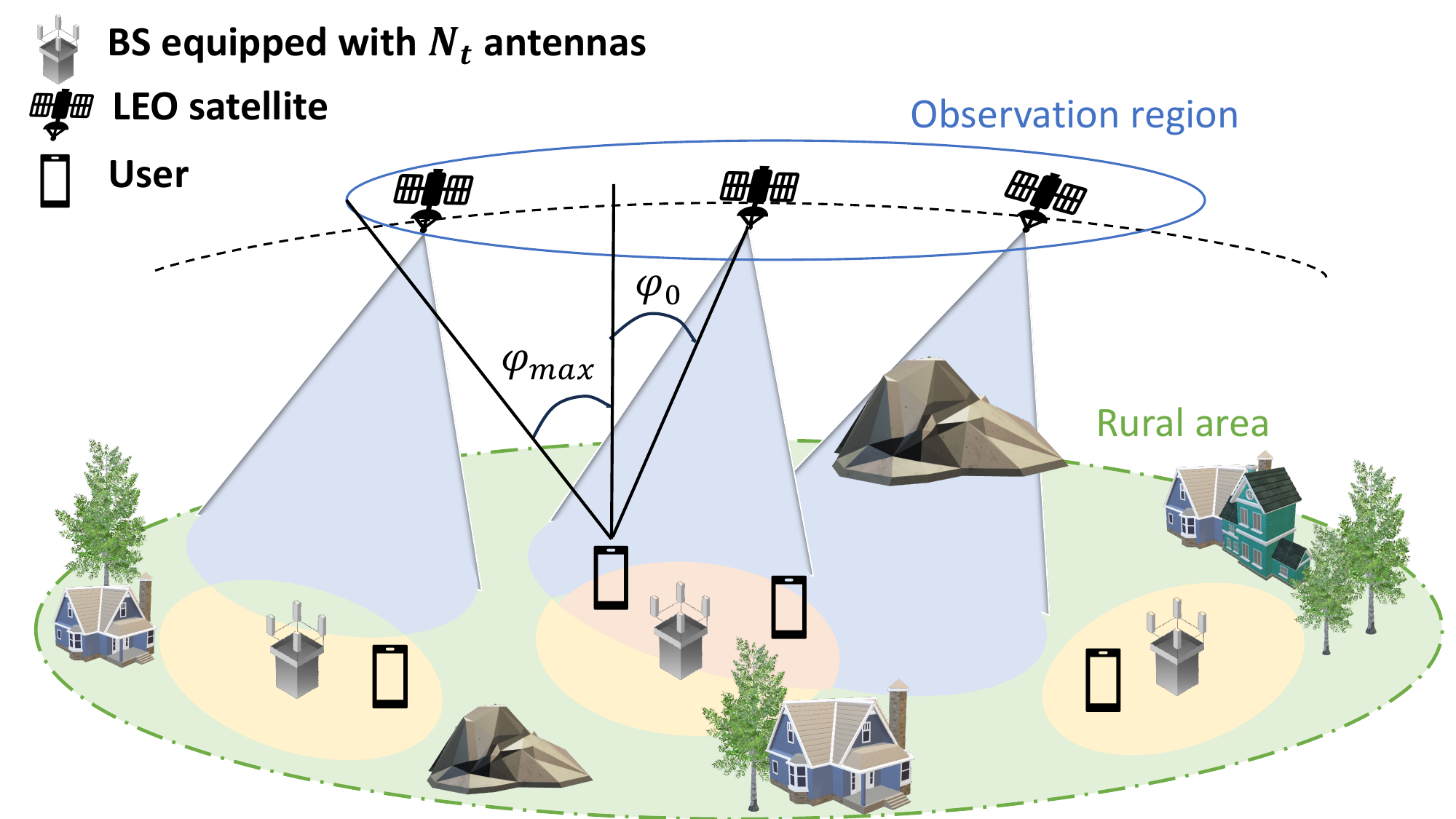}
   \caption{The considered terrestrial/LEO satellite system model.}
   \label{fig:leo}
   \vspace{-0.4cm}
\end{figure}
We consider a 2-tier DL network in which terrestrial BSs and LEO satellites can provide connectivity to single-antenna users in a rural environment. The hybrid terrestrial/satellite network is depicted in Fig.~\ref{fig:leo}. 
\subsection{Terrestrial MIMO Network}
A homogeneous Poisson point process (HPPP) $\Psi_{T}$ with a density $\lambda_{T}$ is used to represent the locations of the terrestrial BSs. We consider a multi-user (MU)-MIMO scenario where each BS transmits with power $P_{t,T}$ and has $N_{T}$ transmit antennas to serve $M$ single-antenna users simultaneously using appropriate precoding schemes. Rayleigh fading is considered to account for small-scale fading. Furthermore, zero-forcing (ZF) is adopted as the implemented precoding technique. 

Without loss of generality, we focus on a typical user located at the origin, and we account for the distance-dependent path loss. Thus, the received power from the serving BS located at distance $r_o$ from the typical user is expressed as 
\begin{equation}
P_{T,o}^r=P_{t,T} r_{o}^{-\eta} \tilde{g}_{o},
\label{eq:16}
\end{equation}
where $\eta$ is the path-loss exponent and $\tilde{g}_{o}$ is the channel parameter which captures the effect of MIMO precoding and is shown to follow a Gamma distribution $\tilde{g}_{o} \sim$ $ \Gamma \left(m_{o}, 1\right)$, $m_{o}=N_T-M+1$ when MU-MIMO with ZF precoding is implemented ~\cite{afify2016unified}. 
\subsection{LEO Satellite Network}
The LEO network constellation comprises $N_{S}$ satellites covering the Earth's surface at a fixed height $h_{S}$. Each satellite has an antenna gain $G_S$ and is transmitting with a transmit power $P_{t,S}$. Following~\cite{al2021analytic}, we model the satellite locations with a HPPP $\Psi_{S}$ of density $\lambda_{S}=\frac{N_{S}}{4\pi\left(R_e+h_{S}\right)^2}$, where $R_e$ is the Earth radius and $4\pi\left(R_e+h_{S}\right)^2$ is the satellite sphere area. As LEO satellites are orbiting the Earth, the user can communicate only with the satellites located in his observation region defined by $\varphi \in\left[0^{\circ}, \varphi_{\max}\right]$, where $\varphi$ is the angle between the center of the Earth pointing to the zenith and the dome edge of last available satellite, $\varphi_{\max}=\cot^{-1}\left(\frac{\alpha}{\sqrt{1-\alpha^2}}\right)$ is the Earth-centered maximum zenith angle and $\alpha=\frac{R_e}{R_e+h_{S}}$~\cite{al2021analytic}. 

To capture the satellite/user line-of-sight (LoS) link, we model the small-scale fading with the Nakagami-m distribution. Thus, the fading gain follows a normalized Gamma distribution with parameter $m$. Furthermore, we account for the free-space distance-dependent path loss $l(\varphi)$ given as
\begin{equation}
\begin{aligned}
l(\varphi)=\frac{l_{\mathrm{o}}}{d^{2}}=\frac{l_{\mathrm{o}}}{R_e^{2}+\left(R_e+h_{S}\right)^{2}-2 R_e\left(R_e+h_{S}\right) \cos \varphi},\\
\end{aligned}
\label{eq:11}
\end{equation}
where $d$ is the distance between the user and the satellite and can be related to the zenith angle $\varphi$ using the cosine rule, $l_{\mathrm{o}}$ is the path-loss constant $\left(l_{\mathrm{o}}=c^{2} /(4 \pi f_S)^{2}\right)$, $c$ is the speed of light and $f_S$ is the carrier frequency of the satellite network. Hence, the received power at the typical user located at the origin of its serving satellite can be formulated as
\begin{equation}
\begin{aligned}
P_{S,o}^r=P_{t,S} G_{S,o} l(\varphi_o) \Omega_o,
\end{aligned}
\label{eq:14}
\end{equation}
where $P_{t,S} $ is the satellite transmit power, $\varphi_o$ is the serving satellite angle, $G_{S,o}$ is the serving satellite main lobe antenna gain and $\Omega_o\sim \Gamma \left(m, \frac{1}{m}\right)$ is the small-scale fading gain. 
\subsection{Association Policy and SINR}\label{sec:association}
We assume that the user is located in a rural environment where terrestrial coverage is not always available. Thus, the satellite network can complement the terrestrial network and provide connectivity. Moreover, we assume that the user associates with the BS or the satellite that provides the highest average received power. This association policy is translated to associating with the nearest BS or the satellite with the lowest contact angle based on the average received power level. The contact angle $\varphi_{\mathrm{o}}$ is defined as the lowest angle between the user pointing to the zenith and an existing satellite in the observation area of the user. The contact angle separates the user from the closest satellite providing the highest average power. The probability density function (PDF) and the cumulative density function (CDF) of $\varphi_{\mathrm{o}}$ are provided in Lemma~\ref{lemma:angle}.
\begin{Lemma}
\normalfont
 The probability density function (PDF) $f_{\varphi_{\mathrm{o}}}(\varphi)$ and  the cumulative distribution function (CDF) $F_{\varphi_{\mathrm{o}}}(\varphi)$ of the contact angle can be expressed as \cite{al2021analytic}
 \begin{equation}
\begin{aligned}
f_{\varphi_{\mathrm{o}}}(\varphi)=\frac{N_{S}}{2} \sin \varphi \exp \left[-\frac{N_{S}}{2}(1-\cos \varphi)\right],
\end{aligned}
\label{eq:7}
\end{equation}
\begin{equation}
\begin{aligned}
F_{\varphi_{\mathrm{o}}}(\varphi)  &\approx 1-\exp \left[-\frac{N_{S}}{2}(1-\cos \varphi)\right].
\end{aligned}
\label{eq:6}
\end{equation} 
\label{lemma:angle}
\vspace{-0.5cm}
\end{Lemma}
We refer to the coverage probability as the main performance metric to evaluate the efficiency of the hybrid terrestrial/satellite network. The coverage probability is defined as the probability that the DL signal-to-interference-and-noise-ratio (SINR) exceeds a predefined threshold $\gamma$. When the user associates with a terrestrial BS, the $\text{SINR}$ is given by 
\begin{equation}
\text{SINR}_{T}=\frac{P_{T,o}^r}{I_{T}+{\sigma}_{T}^2},
\label{eq:17}
\end{equation}
where $P_{T,o}^r$ is the received power from the serving BS given in (\ref{eq:16}), $I_{T}$ is the interference of the terrestrial network, and ${\sigma}_{T}^2$ is the terrestrial noise power. Note here that the terrestrial and the satellite networks are operating at two different frequency bands and are not interfering with each other. Thus, only two types of interference can be encountered in the case of terrestrial association: i) the intra-cell interference caused by other users in the same cell and ii) the inter-cell interference caused by other BSs transmitting on the same carrier frequency. The total terrestrial interference can be expressed as 
\begin{equation}
    \begin{aligned}
     I_{T}=\sum_{r_{i} \in\Psi_{T,o}} P_{t,T} r_{i}^{-\eta} \tilde{g}_{i},
    \end{aligned}
    \label{eq:22}
\end{equation}
where $\Psi_{T,o}$ is the set of interfering BSs, $\tilde{g}_{i}\sim \Gamma \left(m_i=M, 1\right)$ is the equivalent channel gain that captures the effect of MU-MIMO configuration with ZF precoding~\cite{afify2016unified}  and $r_i^{-\eta}$ is the path loss attenuation of the $i^{th}$ interfering BS at a distance $r_i$. Similarly, the DL SINR, when the user associates with a satellite, is given as
\begin{equation}
    \begin{aligned}
       \text{SINR}_{S}=\frac{P_{S,o}^r}{{I}_{S}+\sigma^2_{S}},
    \end{aligned}
\end{equation}
where $P_{S,o}^r$ is the received power from the serving satellite given in (\ref{eq:14}), $\sigma^2_{S}$ is the satellite noise power, and ${I}_{S}$ is the interference from all satellites in the observation area except the serving satellite and is given as
\begin{equation}
\begin{aligned}
I_{S}=\sum_{\varphi_{j} \in \Psi_{S,o}} P_{t,S} G_{S} l(\varphi_j) \Omega_j,
\end{aligned}
\label{eq:18}
\end{equation}
where $\Psi_{S,o}$ is the set of interfering satellites except the serving satellite, $\varphi_j$ is the angle with the $j$-th satellite, $G_S$ is the interfere antennas gain where $G_S < G_{S,o}$ as satellites are interfering with their side lobes, and $\Omega_j$ is the small-scale fading gain of the $j$-the satellite.
\section{Association probabilities and serving distance/angle distributions}
\label{sec:association_ang}
In this section, we provide expressions for the association probabilities with the terrestrial and satellite networks and the distributions of the serving distance and contact angle.
\subsection{Association Probabilities}
In the hybrid terrestrial/satellite network, the user associates with a terrestrial BS that implements the MU-MIMO technique or a satellite according to the maximum average received power. The corresponding association probabilities $A_T$ and $A_S$ are provided in the following lemma. 
\begin{Lemma}[Association probabilities] \normalfont
The probabilities that the user is served by the satellite or the terrestrial networks, denoted by $A_S$ and $A_T$, respectively, are given as
\begin{equation}
    \begin{aligned}
        {A}_{S}=\!\!\int_{0}^{\varphi_{\max}}\!\left[1-F_{R_o}\left(\left(\frac{P_{t,T} m_o}{P_{t,S}G_{S,o}l(\varphi_o)}\right)^{\frac{1}{\eta}}\right)\right] f_{\varphi_{\mathrm{o}}}(\varphi_o) \mathrm{d} \varphi_o,
    \end{aligned}
        \label{A_s}
\end{equation}
\begin{equation}
    \label{A_t}
    \begin{aligned}
      {A}_{T}=1-{A}_{S},
    \end{aligned}
\end{equation}
where $F_{R_o}(r_o)=e^{-\pi \lambda_T r_{o}^2}$ is the CDF of the distance to the nearest terrestrial BS~\cite{hmamouche2021new} and $f_{\varphi_o}(\varphi_o)$ is the satellite contact angle PDF given in (\ref{eq:7}).
\begin{IEEEproof} Following the association rule defined in Section~\ref{sec:association}, the user associates with the nearest BS or the satellite with the lowest contact angle providing the highest average received power.
Thus, the association probability with a satellite denoted as ${A}_{S}$ is derived as
 \begin{equation}
    \begin{aligned}
        {A}_{S}&=\mathbb{P}\left[\bar{P}_{S,o}^r>\bar{P}_{T,o}^r \right] \stackrel{(a)}{=}\mathbb{P}\left[P_{t,S} G_{S,o}l(\varphi_o) >P_{t,T}r_{o}^{-\alpha}m_{o}\right] \\
       &=\mathbb{E}_{\varphi_o}\left[\mathbb{P}\left[r_{o}>\left(\frac{P_{t,T}m_o}{P_{t,S}G_{S,o}l(\varphi_o)}\right)^{\frac{1}{\eta}}\right]\right],
    \end{aligned}
\end{equation}
where (a) is obtained by averaging the received power from the satellite and the terrestrial network given in (\ref{eq:16}) and (\ref{eq:14}) over the channel gain. The remaining of the proof is obtained by referring to the CDF of the distance to the nearest terrestrial BS which is obtained from the void probability of a PPP~\cite{hmamouche2021new}. 
\end{IEEEproof}
\label{lemma:association}
\end{Lemma}
\begin{figure*}
\begin{equation}
    \begin{aligned}
        f_{X_{T}}(x_T)=
        &\begin{cases} 
        \frac{1}{{A}_{T}}f_{R_o}(x_T), & 0\leq x_T\leq  E_T\left(h_S\right),\\
        \frac{1}{{A}_{T}}f_{R_o}(x_T) \!\!\left[1-F_{\varphi_o}\!\!\left(\cos^{-1} \left[\frac{R_e^{2}+\left(R_e+h_{S}\right)^{2}-\frac{P_{t,S}G_{S,o}l_{o}r_o^2}{P_{t,T} m_{o}}}{2R_e\left(R_e+h_{S}\right)}\right]\right)\right], & E_T\left(h_S\right)< x_T\leq E_T\left(\sqrt{R_e^2+\left(R_e+h_S\right)^2}\right), \\
        0, & x_T > E_T\left(\sqrt{R_e^2+\left(R_e+h_S\right)^2}\right),
   \end{cases}
    \end{aligned}
    \label{eq:fXT}
\end{equation}
\hrule
\end{figure*}
\vspace{-0.5cm}
\subsection{Serving Distance/Angle Distributions}
According to the association rule, the user can associate with a terrestrial BS or a satellite. The distributions of the distance to the serving BS and the zenith angle to the serving satellite are provided in Lemma~\ref{lemma:fXT} and Lemma~\ref{lemma:fXS}.
\begin{Lemma}[Terrestrial serving distance distribution] \normalfont 
The conditional distance distribution when the user associates with the terrestrial network is given by (\ref{eq:fXT}) at the top of the next page, where $F_{\varphi_o}(\cdot)$ is the CDF of the satellite contact angle given in (\ref{eq:6}), $E_T(x)=\left(\frac{P_{t,T}m_{o}}{P_{t,S}G_{S,o} l_{o}}\right)^{\frac{1}{\eta}}x^{\frac{2}{\eta}}$, and $f_{R_o}(r_o)=2 \pi \lambda_T r_o e^{-\lambda_T \pi r_o^2}$ is the PDF of the terrestrial contact distance. 
\begin{IEEEproof} The distribution of the distance to the serving BS is equivalent to the distribution of the distance to the nearest BS, given that the user associates with the terrestrial network. Thus, the CDF of $X_T$ can be obtained as 
\begin{equation}
    \begin{aligned}
    &F_{X_T}(x_T)=\mathbb{P}\left[R_o<x|\text{terrestrial association}\right]\\
    &=\frac{\mathbb{P}\left[r_o<x, P_{t,T}r_{o}^{-\alpha}m_{o}>P_{t,S} G_{S,o}l(\varphi_o)\right]}{A_T}\\
    &\stackrel{(a)}{=}\frac{1}{{A}_{T}}\int_{0}^{x_T} f_{R_o}(r_o)\\
    &\mathbb{P}\left[\cos \varphi_o <\frac{R_e^{2}+\left(R_e+h_{S}\right)^{2}-\frac{P_{t,S}G_{S,o}l_{o}r_o^2}{P_{t,T} m_{o}}}{2R_o\left(R_o+h_{S}\right)}\right]\mathrm{d}r_o\\
    &\stackrel{(b)}{=} \begin{cases} 
      F_{X_T}^{(1)}(x_T), & 0\leq x_T\leq E_T\left(h_S\right),\\ 
      F_{X_T}^{(2)}(x_T), & E_T\!\left(h_S\right)\!< x_T \leq \! E_T\!\left(\!\sqrt{R_e^2+\left(R_e+h_S\right)^2}\right), \\
      F_{X_T}^{(3)}(x_T), & x_T > E_T\left(\sqrt{R_e^2+\left(R_e+h_S\right)^2}\right),
   \end{cases}
 \end{aligned}
 \label{eq:CDF_piece}
\end{equation}
where (a) follows from the expression of $l(\varphi_o)$ in (\ref{eq:11}), (b) follows from the fact that $\varphi_o\in\left[0,\frac{\pi}{2}\right]$ and from the CDF definition of $\varphi_o$ and $E_T(x)=\left(\frac{P_{t,T}m_{o}}{P_{t,S}G_{S,o} l_{o}}\right)^{\frac{1}{\eta}}x^{\frac{2}{\eta}}$. The expressions of the piece-wise CDFs of the terrestrial serving distance are obtained from properly integrating the expression in (\ref{eq:CDF_piece}) and are given as
\begin{equation}
    F_{X_T}^{(1)}(x_T)=\frac{1}{{A}_{T}}\int_{0}^{x_T} f_{R_o}(r_o) \mathrm{d}r_o,
\end{equation}
\begin{equation}\small
\begin{aligned}
    &F_{X_{T}}^{(2)}(x_T)=\frac{1}{{A}_{T}}\int_{0}^{E_T\left(h_S\right)}f_{R_o}(r_o)\mathrm{d}r_o + \frac{1}{{A}_{T}}\int_{E_T\left(h_S\right)}^{x_T}f_{R_o}(r_o)\\
    &\left[1-F_{\varphi_o}\left(\cos^{-1} \left[\frac{R_e^{2}+\left(R_e+h_{S}\right)^{2}-\frac{P_{t,S}G_{S,o}l_{o}r_o^2}{P_{t,T} m_{o}}}{2R_e\left(R_e+h_{S}\right)}\right]\right)\right]\mathrm{d}r_o,  
\end{aligned}
\end{equation}
and 
\begin{equation}\small
\begin{aligned}
    &F_{X_T}^{(3)}(x_T)= \frac{1}{{A}_{T}}\int_{0}^{E_T\left(h_S\right)}\!\!\!\!\!\!\!\!\!\!\!\!\!\!f_{R_o}(r_o)\mathrm{d}r_o + \frac{1}{{A}_{T}}\int_{E_T\left(h_S\right)}^{E_T\left(\sqrt{R_e^2+\left(R_e+h_S\right)^2}\right)} \!\!\!\!\!\!\!\!\!\!\!\!\!\!\!\!\!\!\!\!\!\!\!\!\!\!\!\!\!\!\!\!\!\!\!\!\!\!\!\!\!\!f_{R_o}(r_o)\\
    &\left[1-F_{\varphi_o}\left(\cos^{-1} \left[\frac{R_e^{2}+\left(R_e+h_{S}\right)^{2}-\frac{P_{t,S}G_{S,o}l_{o}r_o^2}{P_{t,T} m_{o}}}{2R_e\left(R_e+h_{S}\right)}\right]\right)\right]\mathrm{d}r_o.
\end{aligned}
\end{equation}
The PDF of the serving distance in (\ref{eq:fXT}) is finally obtained by properly deriving the expression in (\ref{eq:CDF_piece}).    
\end{IEEEproof}
\label{lemma:fXT}
\end{Lemma}
\begin{Lemma}[Satellite serving angle distribution] \normalfont The conditional distribution of the contact angle given that the user associates with the satellite network is given as
    \begin{equation}
    \begin{aligned}
        f_{X_{S}}(x_S)=\frac{1}{{A}_{S}}\left[1-F_{R_o}\left(\left(\frac{P_{t,T}m_o}{P_{t,S}G_{S,o}l(x_S)}\right)^{\frac{1}{\eta}}\right)\right] f_{\varphi_0}(x_S),
    \end{aligned}
\end{equation}
where, $F_{R_o}(r_o)=e^{-\pi \lambda_T r_{o}^2}$ is the CDF of the terrestrial contact distance and $f_{\varphi_o}(\cdot)$ is the PDF of the satellite contact angle given in (\ref{eq:7}).
\begin{IEEEproof}
    The proof follows the steps of Lemma~\ref{lemma:fXT}. 
\end{IEEEproof}
\label{lemma:fXS}
\end{Lemma}
\section{Coverage probability}
\label{sec:coverage}
Since the user has two types of association, the total coverage probability of the hybrid terrestrial/satellite system $P_\text{Tot}$ can be obtained using the law of total probability as
\begin{equation}
    \begin{aligned}
        P_\text{Tot}= P_{cov,T} {A}_{T}+ P_{cov,S} {A}_{S},
    \end{aligned}
    \label{eq; cov prob opp}
\end{equation}
where ${A}_{T}$ and ${A}_{S}$ are the association probabilities given in (\ref{A_t}) and (\ref{A_s}). $P_{cov,T}$ and $P_{cov,S}$ are the corresponding conditional coverage probabilities given the association status and are derived in the following theorem.
\begin{Theorem}[Conditional coverage probabilities]\normalfont
    The conditional coverage probabilities given that the user associates with the terrestrial network and the satellite network denoted as $P_{cov,T}$ and $P_{cov,S}$, are given by
    \begin{equation}
    \begin{aligned}
        &P_{cov,T}= \int_{0}^{\infty}f_{X_{T}}(x_{T})\\
        &\sum_{q=0}^{m_{o}-1} \frac{(-s_T(x_T))^{q}}{q!} \left[\frac{\partial^{q}}{\partial_{s_T}^{q}}e^{-s_T(x_T)\sigma_T^2} \mathcal{L}_{I_{T}}(s_{T}(x_{T}))\right]\mathrm{d}x_T,
    \end{aligned}
    \label{eq:PcovT}
    \end{equation}
    where $s_T(x_T)=\frac{\gamma x_T^\eta}{P_{t,T}}$, $f_{X_T}(x_T)$ is given in (\ref{eq:fXT}) and $\mathcal{L}_{I_{T}}(\cdot)$ is the Laplace transform of the terrestrial interference given as
    \begin{equation}
    \begin{aligned}
        &\mathcal{L}_{I_{T}}(s)=\\
        &\exp \left(-\pi\lambda_T x_{T}^2\left[_2F_1\left(\frac{-2}{\eta}, M ; 1-\frac{2}{\eta} ;-s P_{t,T} x_T^{-\eta}\right)-1\right]\right),
    \end{aligned}
    \end{equation}
    and
    \begin{equation}
    \begin{aligned}
        P_{cov,S}=\int_{0}^{\varphi_{\max}} \bar{F}_{\Omega_o}\left(\frac{\gamma \left(\bar{I}_{S}\left(x_S\right)+\sigma^2_S\right)}{P_{t,S} G_{S,o}l\left(x_S\right)}\right) f_{X_{S}}(x_S) \mathrm{d}x_S, \\    
    \end{aligned}
    \end{equation}
    where $\bar{F}_{\mathrm{\Omega_o}}(x)=1-\frac{\Gamma(m,x)}{\Gamma(m)}$ is the complementary CDF (CCDF) of the gamma distribution and $\bar{I}_{S}=\frac{N_S}{2}\int_{x_S}^{\varphi_{\text{max}}}P_{t,S}G_{S}l(\varphi)\sin(\varphi)\mathrm{d}\varphi$ is the average satellite interference.
   \begin{IEEEproof}  
   The conditional coverage probability given that the user associates with a terrestrial BS $P_{cov,T}$ can be derived as
    \begin{equation}
    \begin{aligned}
        &P_{cov,T}= \mathbb{P}\left[\text{SINR}_T\geq\gamma\right]=\mathbb{P}\left[\frac{P_{t,T} x_{T}^{-\eta} \tilde{g}_{o}}{I_{T}+\sigma^2_{T}} \geq\gamma\right]\\
        &=\int_{0}^{\infty}\mathbb{P}\left[ \tilde{g}_{o}\geq\frac{\gamma (I_{T}+\sigma^2_{T})}{P_{t,T} x_{T}^{-\eta}}\right]f_{X_{T}}(x_T)\mathrm{d}x_{T} \\
        &\stackrel{(a)}{=} \int_{0}^{\infty}f_{X_{T}}(x_T)\\
        &\mathbb{E}_{I_{T}}\left[\sum_{q=0}^{m_{o}-1}\frac{1}{q!}\left(\frac{\gamma (I_{T}+\sigma^2_{T})}{P_{t,T} x_{T}^{-\eta}}\right)^{q} \exp \left\{\frac{-\gamma (I_{T}+\sigma^2_{T})}{P_{t,T} x_{T}^{-\eta}}\right\}\right]\mathrm{d}x_{T},\\
    \end{aligned}
    \end{equation}
    where (a) follows from the CDF of the gamma distributed channel parameter $\tilde{g}_{o}$ and from deconditioning over the serving distance distribution $x_T$. The expression in (\ref{eq:PcovT}) is obtained from denoting $s_{T}(x_T)=\frac{\gamma x_T^\eta}{P_{t,T}}$ and from the partial derivative expression of the exponential term and the Laplace transform definition $\mathcal{L}_{I_{T}}(s)=\mathbb{E}_{I_T}\left[e^{-sI_T}\right]$ which can be derived as
    \begin{equation}
    \begin{aligned}
        \mathcal{L}_{I_{T}}(s) & =\mathbb{E}\left[\exp\left(-s\sum_{r_{i} \in\Psi_{T,o}} P_{t,T} r_{i}^{-\eta} \tilde{g}_{i}\right)\right]\\
        &\stackrel{(a)}{=} \exp\left(\!-2\pi\lambda_T\!\!\int_{x_T}^{\infty}\!\left[1-\frac{1}{\left(1+sP_{t,T}r^{-\eta}\right)^{m_i}}\right]\!r\mathrm{d}r\!\right), 
    \end{aligned}
    \end{equation}
    where (a) follows from the probability generating functional (PGFL) of the BSs PPP and from the moment generating functional of the gamma distributed channel gain $\tilde{g}_{i}$ with shape parameter $m_i$ and unity scale parameter. The final expression is obtained by applying the proper change of variables and by solving the integral expression using (p315) from~\cite{gradshteyn2014table}. Finally, the conditional coverage probability $P_{cov,S}$ given that the user associates with a satellite is given as
    \begin{equation}
    \begin{aligned}
        P_{cov,S}&=\mathbb{P}\left[\text{SINR}_S\geq\gamma\right]=\mathbb{P}\left[\frac{P_{t,S} G_{S,o} l(x_S) \Omega_o}{I_{S}+\sigma^2_{S}}>\gamma\right] \\
        &\stackrel{(a)}{\approx}\int_{0}^{\varphi_{\max }}\!\!\mathbb{P}\left[\Omega_o\geq\frac{\gamma \left(\bar{I}_{S}\left(x_S\right)+\sigma^2_{S}\right)}{P_{t,S} G_{S,o} l\left(x_S\right)}\right] f_{X_{S}}(x_S) \mathrm{d}x_S,\\    
    \end{aligned}
    \end{equation}
    where (a) follows from deconditioning over the serving contact angle $x_S$ and from approximating the satellite interference by its average power $\bar{I}_S$. Note that this approximation holds due to the large swath of Earth covered by the satellite which renders the practical number of interfering satellites converge to the mean number. The average satellite interference can be obtained using Campbell's theorem as
    \begin{equation}
    \begin{aligned}
        \bar{I}_S&=\mathbb{E}_{\Omega_j,\Psi_{S,o}}\left[\sum_{\varphi_{j} \in \Psi_{S,o}} P_{t,S} G_{S} l(\varphi_j) \Omega_j\right]\\
        &=\frac{N_S}{2}\int_{x_S}^{\varphi_{\text{max}}}P_{t,S}G_{S}l(\varphi)\sin(\varphi)\mathrm{d}\varphi.
    \end{aligned}
    \end{equation}
 \end{IEEEproof}
 \label{theorem:coverage_prob}
\end{Theorem}

\section{Data Rate}\label{sec:rate}
In this section, we derive the average achievable rate of the hybrid satellite/terrestrial network. The rate expressions are provided in Theorem~\ref{theorem:rate}
\begin{Theorem}[Data Rate] \normalfont
The average achievable rate of the DL hybrid satellite/terrestrial network is given by
\begin{equation}
    \begin{aligned}
        R_\text{Tot}= R_{T} {A}_{T}+R_{S} {A}_{S},
    \end{aligned}
    \label{eq; data rate opp}
\end{equation}
where $A_{T}$ and $A_S$ are given in Lemma~\ref{lemma:association}. $R_{T}$ and $R_{S}$ are the average achievable rates given that the user associates with the terrestrial or the satellite network and are given as
\begin{equation}
    \begin{aligned}
        R_{T}&=\frac{B_T}{\text{ln}2}\int_{0}^{\infty}\!\!\!\frac{1}{1+t}\int_{0}^{\infty} \sum_{q=0}^{m_{o}-1} \frac{1}{q!}\left(\frac{-s_T(x_T)t}{\gamma}\right)^{q} \\
        &\left[\frac{\partial^{q}}{\partial_{s_T}^{q}}e^{-\frac{s_T(x_T)\sigma_T^2 t}{\gamma}}\mathcal{L}_{I_{T}}\left(\frac{s_{T}(x_{T})t}{\gamma}\right)\right]f_{X_{T}}(x_T)\mathrm{d}x_{T}\mathrm{d}t,
    \end{aligned}
\end{equation}
\begin{equation}
    \begin{aligned}
        &R_{S}=\frac{B_S}{\text{ln}2}\\
        &\int_{0}^{\infty}\!\!\!\frac{1}{1+t}\int_{0}^{\varphi_{\max}}  \!\!\!\bar{F}_{\Omega_o}\left(\frac{t \left(\bar{I}_{S}\left(x_S\right)+\sigma^2_S\right)}{P_{t,S} G_{S,o}l\left(x_S\right)}\right) f_{X_S}(x_S) \mathrm{d}x_S \mathrm{d}t,
       \end{aligned}
    \end{equation}
where $s_T(x_T)=\frac{\gamma x_T^\eta}{P_{t,T}}$, ${B}_T$ and ${B}_S$ are the terrestrial and satellite systems bandwidths, $\mathcal{L}_{I_T}(\cdot)$ and $\bar{F}_{\Omega_o}(\cdot)$ are the Laplace transform of terrestrial interference and the CCDF of the serving satellite channel gain and are given in Theorem~\ref{theorem:coverage_prob}.
\begin{IEEEproof}
    The average achievable rate can be derived as
    \begin{equation}
         \begin{aligned}
         &R_\text{Tot}\!\!=\!\mathbb{E}\left[B\text{log}_2\left(1+\text{SINR}\right)\right]\!\!= \!\!\!\int_{0}^{\infty}\!\!\!\!\! \mathbb{P}\left[B\text{log}_2\left(1+\text{SINR}\right)\!>\!y\right]\mathrm{d}y \\
         &\stackrel{(a)}{=}\!\!B_{T}\!\!\!\int_{0}^{\infty}\!\!\!\mathbb{P}\left[\text{SINR}_T>y\right]\mathrm{d}y A_{T}+B_{S}\!\!\!\int_{0}^{\infty}\!\!\!\mathbb{P}\left[\text{SINR}_S>y\right]\mathrm{d}y A_{S},
         \end{aligned}
    \end{equation}
    where (a) is obtained from $\mathbb{E}\left[X\right]=\int_{0}^{\infty}\mathbb{P}\left[X>y\right]\mathrm{d}y$ and from accounting for the two association modes. The remaining of the proof follows a similar methodology as in Theorem~\ref{theorem:coverage_prob} after making the change of variable $t=e^y-1$.
\end{IEEEproof}
    \label{theorem:rate}
\end{Theorem}

 \begin{table}[t!]
\caption{Simulation parameters}
\label{tab:systparam}
\begin{adjustbox}{center}
\begin{tabular}{|c|m{3.4cm}|m{2.6cm}|}
\hline \textbf{Parameter} & \textbf{Description} & \textbf{Value} \\
\hline $N_{S}$, $h_{S}$ & Satellite number and height & $300$~satellites, $800$~km\\
\hline $P_{t,S}$, $\sigma^2_{S}$ &  Satellite transmit and noise powers & $50$~Watt, $-135$~dBm\\
\hline $G_{t}$, $G_{i}$ & Satellite main and side lobe antenna gains & $20$~dB, $5$~dB\\
\hline $N_{T}$, $M$, $\lambda_{T}$ & BS number of antennas, served users and density & $32$ antennas, $16$ users, $5\times 10^{-9}$~BS/m$^2$ \\
\hline $\eta$ & Path-loss exponent & $4$ \\
\hline $P_{t,T}$, $\sigma^2_{T}$ & BS transmit and noise powers & $40$~Watt, $-140$~dBm\\
\hline $f_{S}$, $f_{T}$ & Satellite and BS carrier frequencies & $2$~GHz, $2.5$~GHz\\
\hline $B_{S}$, $B_{T}$ & Satellite and BS bandwidths & $200$~MHz, $50$~MHz\\
\hline
\end{tabular}
\end{adjustbox}
\end{table} 
\begin{figure}[t!]
    \centering
    \includegraphics[width=1.05\linewidth]{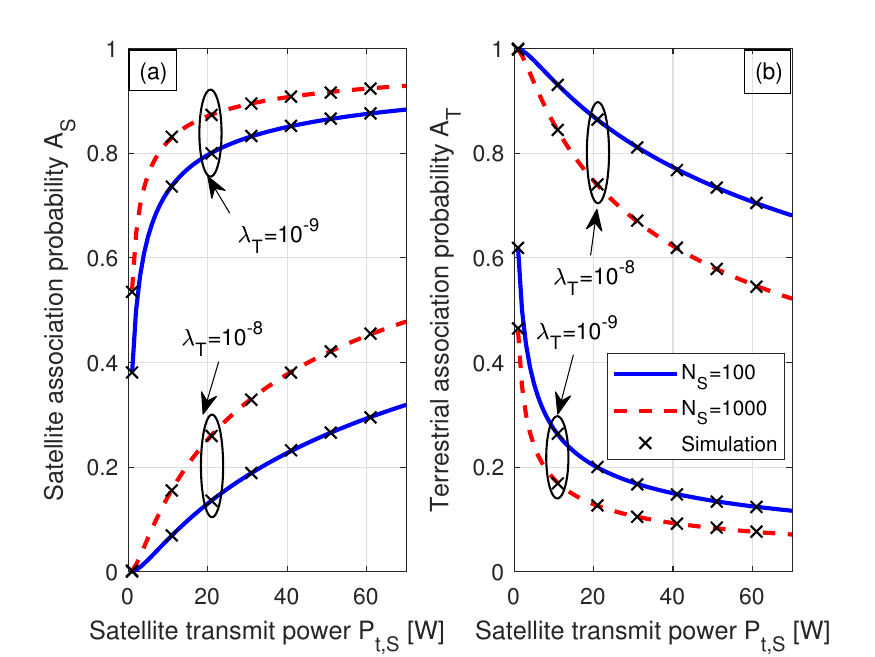}
    \caption{Association probabilities as a function of the satellite transmit power $P_{t,S}$ with different $\lambda_T$ and $N_S$.}
    \label{ass prob}
\end{figure}
\section{Numerical results}
\label{sec:resutls}
In this section, we study the performance of the hybrid terrestrial/satellite system in a rural environment with sparse BS distribution and conduct intensive Monte Carlo simulations to validate the analytical results. The simulation parameters are given in Table~\ref{tab:systparam}. Note that the simulation parameters for the LEO network are inspired from~\cite{al2021analytic} and for the terrestrial network from typical values in rural areas \cite{3GPPNR}. In Fig.~\ref{ass prob}, we highlight the effect of the terrestrial BSs density and number of satellites on the association probabilities $A_S$ and $A_T$ while changing the satellite transmit power $P_{t,S}$. Fig.~\ref{ass prob} shows that analytical and simulation results match, validating our derived equations. Moreover, a higher satellite transmit power or a larger number of satellites increases the chance of having a nearby satellite that provides the highest average received power, thus, increasing the satellite association probability. On the other hand, a relatively denser terrestrial network ($\lambda_T=10^{-8}$ BS/m$^2$) limits the association with the satellite network as the user can still find a good nearby BS. However, in rural areas with limited density of BS ($\lambda_T=10^{-9}$ BS/m$^2$), the satellite network is more likely the only option to provide connectivity to the user, and the satellite association probability increases. 
\begin{figure}[t!]
\centering
\resizebox{0.98\columnwidth}{!}{\begin{tikzpicture}[thick,scale=1, every node/.style={scale=1.3},font=\Huge]
%
%

\begin{axis}[%
width=9.2in,
height=7.9in,
at={(0.882in,0.44in)},
scale only axis,
xmin=1,
xmax=15,
xtick={1, 5, 10, 15},
xlabel style={font=\color{white!15!black}},
xlabel={\Huge{Number of simultaneously connected users M}},
ymin=0.3,
ymax=1,
ytick={0.3,0.4,...,1},
ylabel style={font=\color{white!15!black}},
ylabel={\Huge{Coverage probability}},
axis background/.style={fill=white},
xmajorgrids,
ymajorgrids,
legend style={at={(0.05,0.05)}, anchor=south west, legend cell align=left, align=left, draw=white!15!black}
]
\addplot [color=blue, dashed, line width=3.5pt]
  table[row sep=crcr]{%
1	0.92963\\
3	0.8454\\
5	0.73698\\
7	0.64036\\
9	0.55979\\
11	0.48742\\
13	0.43074\\
15	0.38241\\
};
\addlegendentry{$\text{N}_\text{T}\text{=32}$, Terrestrial}

\addplot [color=red, dashed, line width=3.5pt]
  table[row sep=crcr]{%
1	0.9808\\
3	0.95397\\
5	0.91247\\
7	0.85787\\
9	0.79982\\
11	0.74187\\
13	0.68451\\
15	0.63221\\
};
\addlegendentry{$\text{N}_\text{T}\text{=64}$, Terrestrial}

\addplot [color=blue, line width=3.5pt]
  table[row sep=crcr]{%
1	0.87631\\
3	0.87309\\
5	0.85954\\
7	0.8397\\
9	0.81723\\
11	0.7918\\
13	0.76891\\
15	0.7536\\
};
\addlegendentry{$\text{N}_\text{T}\text{=32}$, Hybrid}

\addplot [color=red, line width=3.5pt]
  table[row sep=crcr]{%
1	0.90937\\
3	0.90924\\
5	0.90692\\
7	0.90055\\
9	0.88947\\
11	0.87766\\
13	0.85676\\
15	0.83657\\
};
\addlegendentry{$\text{N}_\text{T}\text{=64}$, Hybrid}

\addplot [color=black, line width=2.0pt, only marks, mark size=10.0pt, mark=x, mark options={solid, black}]
  table[row sep=crcr]{%
1	0.92963\\
3	0.8454\\
5	0.73698\\
7	0.64036\\
9	0.55979\\
11	0.48742\\
13	0.43074\\
15	0.38241\\
};
\addlegendentry{Simulation}

\addplot [color=black, line width=2.0pt, only marks, mark size=10.0pt, mark=x, mark options={solid, black}, forget plot]
  table[row sep=crcr]{%
1	0.9808\\
3	0.95397\\
5	0.91247\\
7	0.85787\\
9	0.79982\\
11	0.74187\\
13	0.68451\\
15	0.63221\\
};

\addplot [color=black, line width=2.0pt, only marks, mark size=10.0pt, mark=x, mark options={solid, black}, forget plot]
  table[row sep=crcr]{%
1	0.90937\\
3	0.90924\\
5	0.90692\\
7	0.90055\\
9	0.88947\\
11	0.87766\\
13	0.85676\\
15	0.83657\\
};

\addplot [color=black, line width=2.0pt, only marks, mark size=10.0pt, mark=x, mark options={solid, black}, forget plot]
  table[row sep=crcr]{%
1	0.87631\\
3	0.87309\\
5	0.85954\\
7	0.8397\\
9	0.81723\\
11	0.7918\\
13	0.76891\\
15	0.7536\\
};

\end{axis}
\end{tikzpicture}}
    \caption{Coverage probability as a function of the number of users $M$ for different numbers of BS transmit antennas $N_T$.}
    \label{cvg prob M_N}
\end{figure}
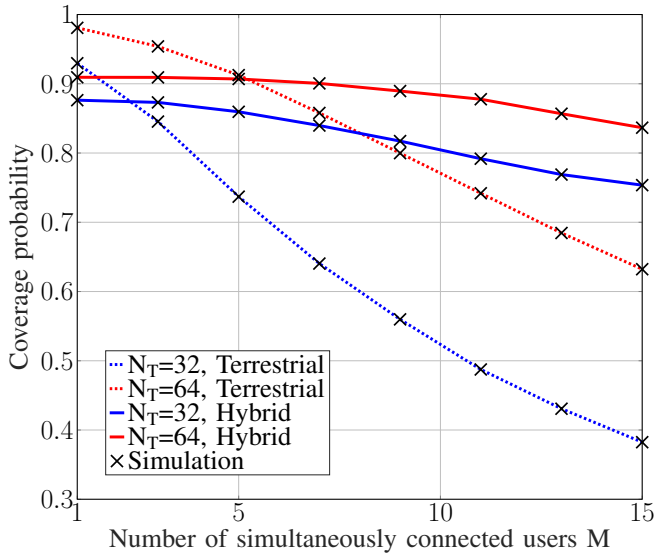

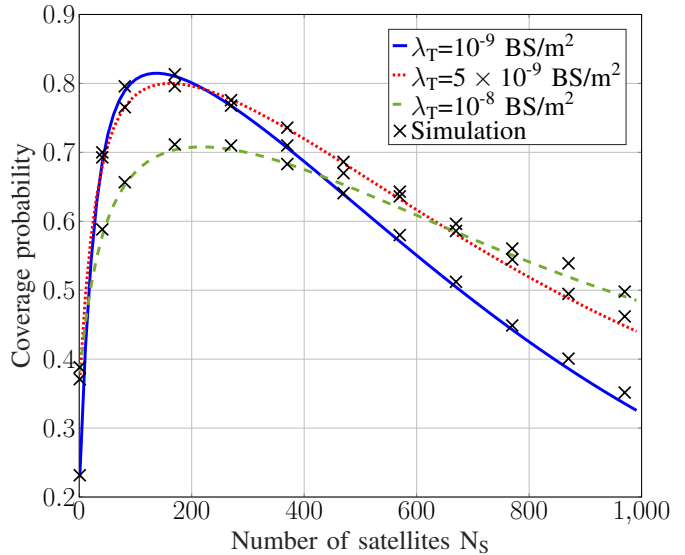
\begin{figure}[t!]
    \centering
    \resizebox{\columnwidth}{!}
    {\begin{tikzpicture}[thick,scale=1, every node/.style={scale=1.3},font=\Huge]
%
%
\definecolor{mycolor1}{rgb}{0.46667,0.67451,0.18824}%

\begin{axis}[%
width=9.2in,
height=7.9in,
at={(0.882in,0.566in)},
scale only axis,
xmin=0,
xmax=1000,
xtick={0,200,...,1000},
xlabel style={font=\color{white!15!black}},
xlabel={\Huge{$\text{Number of satellites N}_\text{S}$}},
ymin=0.2,
ymax=0.9,
ytick={0.2,0.3,...,0.9},
ylabel style={font=\color{white!15!black}},
ylabel={\Huge{Coverage probability}},
axis background/.style={fill=white},
xmajorgrids,
ymajorgrids,
legend style={at={(0.55,0.73)}, anchor=south west, legend cell align=left, align=left, draw=white!15!black}
]
\addplot [color=blue, line width=3.5pt]
  table[row sep=crcr]{%
1	0.228247468322428\\
11	0.415163744655025\\
21	0.541143471365697\\
31	0.626663988255993\\
41	0.685370177154322\\
51	0.726146648236645\\
61	0.754758011247282\\
71	0.774964977448095\\
81	0.789246767914348\\
91	0.799260268539864\\
101	0.806130469010449\\
111	0.810634560317834\\
121	0.813319347529125\\
131	0.814576803174668\\
141	0.814693218969833\\
151	0.813881582578688\\
161	0.812303196652315\\
171	0.810082324125224\\
181	0.807316258212697\\
191	0.804082351636479\\
201	0.800442997299881\\
211	0.796449209432084\\
221	0.792143234990933\\
231	0.787560483623789\\
241	0.782730972145816\\
251	0.777680418517705\\
261	0.772431079541879\\
271	0.767002398902955\\
281	0.761411513262682\\
291	0.755673650989248\\
301	0.749802448873378\\
311	0.743810205621251\\
321	0.737708086193163\\
331	0.731506287623146\\
341	0.725214174431266\\
351	0.71884038986736\\
361	0.712392947822238\\
371	0.70587930918231\\
381	0.699306445596211\\
391	0.692680893002242\\
401	0.686008796786181\\
411	0.679295950066125\\
421	0.672547826308723\\
431	0.665769607250939\\
441	0.658966206918947\\
451	0.652142292390238\\
461	0.645302301828617\\
471	0.638450460227841\\
481	0.631590793223971\\
491	0.624727139274755\\
501	0.617863160454321\\
511	0.611002352070118\\
521	0.604148051275417\\
531	0.597303444822715\\
541	0.59047157608029\\
551	0.583655351415087\\
561	0.576857546029041\\
571	0.570080809322673\\
581	0.563327669848657\\
591	0.556600539908608\\
601	0.549901719838475\\
611	0.543233402021301\\
621	0.536597674660401\\
631	0.529996525341319\\
641	0.523431844406798\\
651	0.516905428165596\\
661	0.510418981953048\\
671	0.503974123058707\\
681	0.497572383534386\\
691	0.491215212893884\\
701	0.484903980714382\\
711	0.478639979147899\\
721	0.472424425350164\\
731	0.466258463833359\\
741	0.460143168748078\\
751	0.454079546099404\\
761	0.448068535901218\\
771	0.442111014272302\\
781	0.436207795477432\\
791	0.430359633916181\\
801	0.424567226061847\\
811	0.418831212352579\\
821	0.413152179036599\\
831	0.407530659973111\\
841	0.401967138390401\\
851	0.396462048602357\\
861	0.391015777684611\\
871	0.385628667111334\\
881	0.380301014353612\\
891	0.37503307444026\\
901	0.369825061481885\\
911	0.364677150158895\\
921	0.359589477174143\\
931	0.354562142670858\\
941	0.349595211616401\\
951	0.344688715152511\\
961	0.339842651912494\\
971	0.335056989305925\\
981	0.330331664771381\\
991	0.32566658699767\\
};
\addlegendentry{$\lambda{}_\text{T}\text{=10}^{\text{-9}}\text{ BS/m}^\text{2}$}

\addplot [color=red, dashed, line width=3.5pt]
  table[row sep=crcr]{%
1	0.369439975307134\\
11	0.49186137308005\\
21	0.576614285484099\\
31	0.636480472707207\\
41	0.67958420112498\\
51	0.711161185749215\\
61	0.734637460143785\\
71	0.752292297992596\\
81	0.765669048808935\\
91	0.775832216777788\\
101	0.78353006640625\\
111	0.78929869681348\\
121	0.793529514532522\\
131	0.796513582339741\\
141	0.798471194328525\\
151	0.799571897354343\\
161	0.799948254489006\\
171	0.799705453479517\\
181	0.798928117640638\\
191	0.797685206080875\\
201	0.796033590148842\\
211	0.794020699591446\\
221	0.791686505799292\\
231	0.789065026276334\\
241	0.786185478846677\\
251	0.783073176469419\\
261	0.77975022773473\\
271	0.776236090210902\\
281	0.772548011233101\\
291	0.768701381780282\\
301	0.764710022653004\\
311	0.760586417485944\\
321	0.756341903690132\\
331	0.751986829867511\\
341	0.747530686328093\\
351	0.742982213894409\\
361	0.738349495076282\\
371	0.733640030852643\\
381	0.728860805642201\\
391	0.724018342534579\\
401	0.71911875045318\\
411	0.714167764605339\\
421	0.709170781324446\\
431	0.704132888208679\\
441	0.699058890300356\\
451	0.693953332920337\\
461	0.688820521667053\\
471	0.683664540004178\\
481	0.678489264791226\\
491	0.6732983800539\\
501	0.668095389243919\\
511	0.662883626198765\\
521	0.65766626497952\\
531	0.652446328737807\\
541	0.647226697740315\\
551	0.642010116660434\\
561	0.636799201230585\\
571	0.631596444335409\\
581	0.626404221614649\\
591	0.621224796634906\\
601	0.616060325681249\\
611	0.610912862212784\\
621	0.605784361020211\\
631	0.600676682118454\\
641	0.595591594402954\\
651	0.590530779094574\\
661	0.585495832994805\\
671	0.580488271570152\\
681	0.575509531882294\\
691	0.570560975378352\\
701	0.565643890554056\\
711	0.560759495500807\\
721	0.555908940346423\\
731	0.551093309598202\\
741	0.546313624395762\\
751	0.541570844680424\\
761	0.536865871287011\\
771	0.532199547963267\\
781	0.527572663321556\\
791	0.522985952726969\\
801	0.518440100125501\\
811	0.513935739815568\\
821	0.509473458165818\\
831	0.505053795281844\\
841	0.500677246624207\\
851	0.496344264579838\\
861	0.492055259988817\\
871	0.487810603628249\\
881	0.483610627654835\\
891	0.479455627007592\\
901	0.475345860772089\\
911	0.471281553507392\\
921	0.467262896536884\\
931	0.463290049203998\\
941	0.459363140093874\\
951	0.455482268221832\\
961	0.451647504189541\\
971	0.447858891309689\\
981	0.444116446699953\\
991	0.44042016234697\\
};
\addlegendentry{$\lambda{}_\text{T}\text{=5}\times\text{10}^{\text{-9}}\text{ BS/m}^\text{2}$}

\addplot [color=mycolor1, dashed, dash pattern=on 10pt off 10pt on 10pt off 10pt, line width=3.5pt]
  table[row sep=crcr]{%
1	0.38133864421165\\
11	0.453075638688636\\
21	0.506408970471744\\
31	0.546943015581548\\
41	0.578372403847657\\
51	0.603175387340486\\
61	0.62304513744293\\
71	0.639160762883946\\
81	0.652359533106729\\
91	0.663247671702306\\
101	0.6722725823389\\
111	0.679770602826476\\
121	0.685999051365014\\
131	0.691158064790661\\
141	0.695405715243028\\
151	0.698868639620396\\
161	0.701649630589811\\
171	0.703833140193724\\
181	0.705489328451707\\
191	0.706677083086357\\
201	0.70744630140944\\
211	0.707839635847036\\
221	0.707893844467058\\
231	0.707640847006203\\
241	0.707108558757309\\
251	0.706321555061432\\
261	0.705301605301881\\
271	0.704068105405045\\
281	0.702638430703403\\
291	0.701028225791487\\
301	0.699251644147098\\
311	0.697321547411958\\
321	0.695249672058737\\
331	0.693046769524849\\
341	0.690722724632044\\
351	0.688286656136619\\
361	0.685747002497413\\
371	0.683111595354952\\
381	0.680387722746929\\
391	0.6775821837137\\
401	0.67470133565075\\
411	0.671751135527047\\
421	0.668737175895948\\
431	0.665664716469454\\
441	0.662538711899541\\
451	0.659363836306202\\
461	0.656144505006317\\
471	0.652884893826748\\
481	0.649588956326493\\
491	0.646260439203926\\
501	0.642902896124484\\
511	0.639519700169911\\
521	0.636114055081553\\
531	0.632689005445906\\
541	0.629247445950136\\
551	0.625792129817913\\
561	0.622325676521035\\
571	0.618850578849752\\
581	0.61536920941382\\
591	0.61188382663713\\
601	0.608396580300665\\
611	0.604909516681812\\
621	0.601424583332001\\
631	0.597943633529625\\
641	0.594468430440637\\
651	0.591000651015469\\
661	0.587541889647464\\
671	0.584093661615151\\
681	0.580657406328094\\
691	0.577234490393817\\
701	0.573826210521381\\
711	0.570433796275416\\
721	0.567058412692932\\
731	0.563701162773948\\
741	0.560363089855692\\
751	0.557045179879218\\
761	0.553748363556325\\
771	0.550473518443821\\
781	0.547221470931512\\
791	0.543992998149697\\
801	0.540788829801271\\
811	0.537609649923153\\
821	0.534456098581291\\
831	0.531328773503047\\
841	0.528228231650499\\
851	0.525154990737801\\
861	0.522109530695538\\
871	0.519092295084712\\
881	0.516103692462803\\
891	0.513144097704118\\
901	0.510213853276523\\
911	0.507313270476405\\
921	0.504442630623659\\
931	0.501602186218311\\
941	0.498792162060259\\
951	0.496012756333592\\
961	0.493264141656745\\
971	0.490546466099738\\
981	0.48785985416966\\
991	0.485204407765459\\
};
\addlegendentry{$\lambda{}_\text{T}\text{=10}^{\text{-8}}\text{ BS/m}^\text{2}$}

\addplot [color=black, line width=2.0pt, only marks, mark size=10.0pt, mark=x, mark repeat={2}, mark options={solid, black}]
  table[row sep=crcr]{%
1	0.2316\\
21	0.5564\\
41	0.7\\
61	0.7641\\
81	0.7958\\
120	0.8115\\
170	0.8132\\
220	0.7853\\
270	0.7678\\
320	0.7299\\
370	0.71\\
420	0.6756\\
470	0.6408\\
520	0.6075\\
570	0.58\\
620	0.5451\\
670	0.5122\\
720	0.4784\\
770	0.4489\\
820	0.4226\\
870	0.4008\\
920	0.3832\\
970	0.3515\\
};
\addlegendentry{Simulation}

\addplot [color=black, line width=2.0pt, only marks, mark size=10.0pt, mark=x, mark repeat={2}, mark options={solid, black}, forget plot]
  table[row sep=crcr]{%
1	0.3709\\
21	0.5868\\
41	0.6927\\
61	0.7412\\
81	0.7654\\
120	0.7971\\
170	0.7962\\
220	0.7929\\
270	0.7757\\
320	0.7551\\
370	0.7358\\
420	0.7104\\
470	0.6862\\
520	0.6679\\
570	0.643\\
620	0.6125\\
670	0.586\\
720	0.5655\\
770	0.5446\\
820	0.5193\\
870	0.4947\\
920	0.4825\\
970	0.462\\
};
\addplot [color=black, line width=2.0pt, only marks, mark size=10.0pt, mark=x, mark repeat={2}, mark options={solid, black}, forget plot]
  table[row sep=crcr]{%
1	0.3882\\
21	0.5208\\
41	0.5882\\
61	0.6316\\
81	0.6566\\
120	0.6975\\
170	0.7114\\
220	0.7147\\
270	0.7099\\
320	0.6946\\
370	0.683\\
420	0.6671\\
470	0.6698\\
520	0.6609\\
570	0.6361\\
620	0.6151\\
670	0.5963\\
720	0.5752\\
770	0.5605\\
820	0.5444\\
870	0.539\\
920	0.5153\\
970	0.4979\\
};
\end{axis}
    \end{tikzpicture}}
    \caption{Hybrid system coverage probability as a function of the number of satellites $N_S$ and the BSs density $\lambda_T$.}
    \label{cvg prob nsat}
\end{figure}

Fig.~\ref{cvg prob M_N} plots the coverage probability as a function of the number of simultaneously served users $M$ by the MU-MIMO BSs, and the number of BSs transmit antennas. Furthermore, Fig.~\ref{cvg prob M_N} compares the performance of the hybrid terrestrial/satellite system with the terrestrial-only system when the satellite network is ignored. Fig.~\ref{cvg prob M_N} shows that adding more antennas improves the coverage probability, whereas connecting more users simultaneously leads to a decrease in the coverage probability due to intra-cell interference. Furthermore, for a low number of users, the coverage probability of the terrestrial network exceeds the hybrid terrestrial/satellite system, which starts outperforming the terrestrial network at a certain number of users $M$ function of the number of transmit antennas. This is mainly due to the fact that, for a lower number of simultaneously served users, the intra-cell interference in the terrestrial network is dominated by interference from the satellite network. Thus, the hybrid terrestrial/satellite system does not provide any potential gains. On the other side, as the number of users in a cell increases, the terrestrial network will no longer be able to cope with the increased intra-cell interference levels, and the satellite network can help offload the traffic to improve the coverage probability. Moreover, having more BS transmit antennas improves the capability of the terrestrial network to accommodate more users per cell while alleviating the need for the satellite network. 

The impact of the number of satellites $N_S$ and the density of the terrestrial BSs on the total coverage probability is depicted in Fig.~\ref{cvg prob nsat}. We can note that as the number of satellites increases, the coverage probability improves. This happens up to a certain optimal number of satellites, after which the coverage probability starts to deteriorate. A higher number of satellites increases the association probability with the satellite network and improves the connection with the serving satellite. However, the satellite interference rises as well, and after a certain threshold, dominates the improvement of the useful signal power, resulting in a low SINR and a degradation of the coverage probability. Fig.~\ref{cvg prob nsat} also shows that the coverage probability performance in a rural environment with a low density of BSs outperforms the case of a more dense setup if the number of deployed satellites is low. However, for dense satellite deployments, such behavior flips, and a denser terrestrial setup ($\lambda_T=10^{-8}$~BS/m$^2$) shows an improved coverage probability compared with a sparser deployment ($\lambda_T=10^{-9}$~BS/m$^2$). Such behavior is quite pronounced since, with higher BS densities, the user associates more with the terrestrial network and is, therefore, less affected by satellite interference. However, as the density of BSs decreases, the user is forced to switch to the satellite tier, and its performance is mainly defined by the satellite interference levels.

\begin{figure}[t!]
    \centering\resizebox{\columnwidth}{!}
    {\begin{tikzpicture}[thick,scale=1, every node/.style={scale=1.3},font=\Huge]
%
%
\definecolor{mycolor1}{rgb}{0.46667,0.67451,0.18824}%

\begin{axis}[%
width=9.2in,
height=7.9in,
at={(0.882in,0.61in)},
scale only axis,
xmin=0,
xmax=1000,
xtick={0,200,...,1000},
xlabel style={font=\color{white!15!black}},
xlabel={\Huge{$\text{Number of satellites N}_\text{S}$}},
ymin=0,
ymax=600,
ytick={0,100,...,600},
ylabel style={font=\color{white!15!black}},
ylabel={\Huge{Data rate [Mbps]}},
axis background/.style={fill=white},
xmajorgrids,
ymajorgrids,
legend style={at={(0.54,0.03)}, anchor=south west, legend cell align=left, align=left, draw=white!15!black}
]
\addplot [color=blue, line width=3.5pt, mark=o, mark size=8.0pt, mark repeat={2}, mark options={solid, blue}]
  table[row sep=crcr]{%
0	70.8185564244104\\
50	493.810513405853\\
100	553.312587571652\\
150	552.006505951443\\
200	540.330512457457\\
250	523.768906344496\\
300	505.47915170504\\
350	486.662286756047\\
400	472.416591073905\\
450	459.227449214576\\
500	446.763128539025\\
550	431.104895081174\\
600	418.038359525717\\
650	407.502389427909\\
700	401.826808507067\\
750	388.053857442355\\
800	378.277663063174\\
850	369.282913919848\\
900	362.65214740126\\
950	355.275294048033\\
1000	347.970351288513\\
};
\addlegendentry{$\lambda{}_\text{T}\text{=10}^{\text{-9}}\text{ BS/m}^\text{2}$}

\addplot [color=red, line width=3.5pt, mark=asterisk, mark size=8.0pt, mark repeat={2}, mark options={solid, red}]
  table[row sep=crcr]{%
0	113.457599331404\\
50	387.478030584583\\
100	438.97979171987\\
150	447.679064335143\\
200	446.299319882517\\
250	440.261827817121\\
300	429.735976085771\\
350	420.979933403172\\
400	412.245336706883\\
450	397.762738247824\\
500	389.681815333828\\
550	383.898649354826\\
600	375.292762069968\\
650	363.202404976346\\
700	357.799655233006\\
750	350.235316390788\\
800	343.678795561095\\
850	339.071969563791\\
900	332.729101787568\\
950	324.165452524079\\
1000	319.007823266088\\
};
\addlegendentry{$\lambda{}_\text{T}\text{=5}\times\text{10}^{\text{-9}}\text{ BS/m}^\text{2}$}

\addplot [color=mycolor1, line width=3.5pt, mark=triangle, mark size=8.0pt, mark repeat={2}, mark options={solid, mycolor1}]
  table[row sep=crcr]{%
0	119.17980202041\\
50	284.157175685872\\
100	331.433106652893\\
150	346.13027621348\\
200	346.682324978235\\
250	345.535587519965\\
300	343.707028759544\\
350	341.055721299038\\
400	334.953375231998\\
450	328.961360316665\\
500	320.645531847581\\
550	319.472693511225\\
600	311.561757320581\\
650	306.505369448059\\
700	301.510780353324\\
750	298.72845604562\\
800	292.026772872257\\
850	287.535276143291\\
900	284.442679738442\\
950	279.562894992918\\
1000	273.635866735204\\
};
\addlegendentry{$\lambda{}_\text{T}\text{=10}^{\text{-8}}\text{ BS/m}^\text{2}$}

\end{axis}
    \end{tikzpicture}}
    \caption{Hybrid system data rate as a function of the number of satellites $N_S$ and the BSs density $\lambda_T$.}
    \label{data rate nsat}
\end{figure}
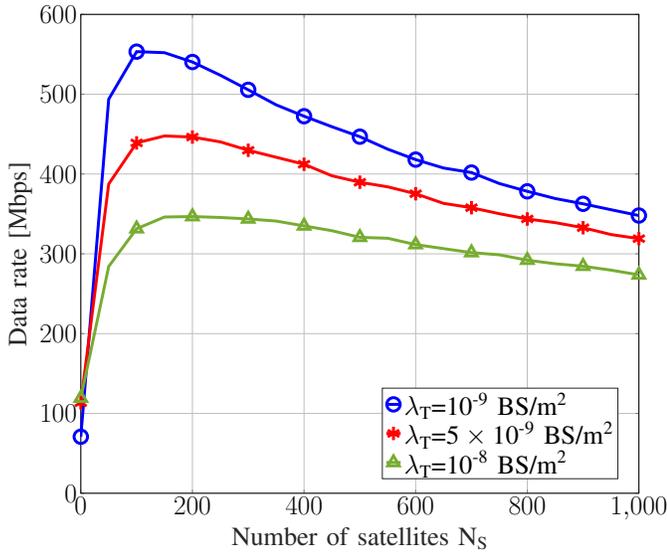

Finally, Fig.~\ref{data rate nsat} illustrates the impact of the number of satellites $N_S$ and the density of terrestrial BSs $\lambda_T$ on the average achievable rate of the hybrid terrestrial/satellite system. Similarly to the coverage probability performance in Fig.~\ref{cvg prob nsat}, adding more satellites increases the total data rate up to a certain number of satellites beyond which the data rate deteriorates due to increased interference levels. An optimal number of satellites must therefore be deployed to maximize the data rate. However, we can note that the hybrid terrestrial/satellite system can still ensure a better data rate compared to the terrestrial system ($N_S=0$) irrespective of the number of added satellites. Furthermore, when compared to the coverage probability in Fig.~\ref{cvg prob nsat}, a clear behavior mismatch can be noticed as the data rate always improves with the decrease of the BSs density. In fact, for low-density of BSs, the user is forced to associate more with the satellite network and can therefore benefit from the higher offered bandwidth.  

\section{Conclusion}
\label{sec:conclusion}
In this paper, we studied the performance of a hybrid terrestrial/satellite system in providing connectivity to rural areas. Using tools from stochastic geometry, we derived expressions for the association probabilities with each tier, the coverage probability, and the average achievable rate. We characterized the impact of both the satellite network and the terrestrial network in terms of the constellation size and transmit power of satellites and the density and MIMO configuration of terrestrial BSs. The obtained results highlight the superiority of the hybrid system compared to the terrestrial network and provide insights to guide the design of hybrid terrestrial/satellite infrastructures for a desired quality of service in any type of environment. This is of high interest as ``connecting the unconnected" is one of the main goals of 6G systems.   
\bibliographystyle{IEEEtran}
\bibliography{ref}
\end{document}